\documentclass[12pt]{article} 

\usepackage{cancel}

\usepackage{citesort} 

\usepackage{undertilde}

\usepackage{graphicx}

\title{Does a Second-Class Primary Constraint Generate a Gauge Transformation?  Electromagnetisms and Gravities, Massless and Massive\\Accepted on 9 February 2024 to \emph{Annals of Physics} }

  \author{J. Brian Pitts \\University of Cambridge, University of Lincoln, and University of South Carolina \\ Supported by National Science Foundation (USA) grant \#1734402 and \\John Templeton Foundation grant \#60745 \\ ORCID \#0000-0002-7299-5137}  

 \date{}

\begin{document}

\maketitle

\begin{abstract} 

In constrained Hamiltonian dynamics there are two views regarding how  first-class constraints generate  gauge transformations:  individually or only  in a certain combination, the Rosenfeld-Anderson-Bergmann-Castellani gauge generator.  This gauge generator $G$ preserves Hamilton's equations and changes the canonical action at most  by a boundary term; Hamiltonian's equations are the Euler-Lagrange equations for the canonical action.  Hence the canonical formalism is equivalent to the Lagrangian formalism, and indeed subsumed within it (in important examples) with many canonical momenta serving as auxiliary fields.  $G$ generates transformations basically equivalent to the usual 4-dimensional Lagrangian expressions, such as a 4-gradient in electromagnetism or a space-time coordinate transformation (on-shell) in General Relativity.  It has been shown recently that separate first-class constraints lead to inequivalent observables between Proca non-gauge and Stueckelberg gauge massive electromagnetism.  There is, however, widespread agreement that second-class constraints do not generate gauge transformations. 

Here it is shown that in such a sense as the first-class primary constraint in Maxwell's theory generates a gauge transformation, the second-class primary constraint in Proca's massive electromagnetism also generates a gauge transformation. Likewise the second-class primary constraints in various massive spin $2$ relatives of General Relativity generate as much of a gauge transformation as do the corresponding first-class primary constraints in GR.  Hence the view that first-class constraints typically generate gauge transformations \emph{individually} faces a puzzle not faced by the gauge generator view.

\end{abstract}

keywords:  gauge freedom, observables, constrained Hamiltonian dynamics, constraints


\section{Gauge Freedoms in Electromagnetisms? } 

The relationship between the Lagrangian and Hamiltonian formalisms continues to be more tense than one might expect given the undergraduate derivation of the Hamiltonian formalism from the Lagrangian.  This tension exists at the level of constrained systems, in which the Legendre transformation cannot be performed fully.  Constrained systems includes theories with gauge freedom such as electromagnetism and Yang-Mills and (in a broader sense) General Relativity, where all the constraints (in the most usual formulation) are ``first-class,'' which implies having vanishing Poisson brackets among themselves (on-shell if needed).  Spinor fields are also constrained due to being only linear in time derivatives; the resulting constraints are ``second-class'' (having non-zero Poisson brackets among themselves) and do not reflect gauge freedom.  Hence the  bulk of the fields relevant in fundamental physics are constrained \cite{Sundermeyer}.  This tension between the Lagrangian and Hamiltonian formalisms bears upon issues in quantum gravity such as the problem of time, including questions such as $3$-dimensional \emph{vs.} $4$-dimensional covariance \cite{IshamPrima,KucharCanonical93,PonsSalisburyShepleyYang,PonsSalisburySundermeyerFolklore,GRChangeNoKilling}.

There is an angle from which the Hamiltonian formalism, even if constrained, can hardly fail to be equivalent to the Lagrangian classically.  Hamiltonian's equations are the Euler-Lagrange equations for the canonical action $\int  (p \dot{q} -H)dt$.  Hence the canonical formalism is equivalent to the Lagrangian formalism, and indeed subsumed within it (at least in many important examples including electromagnetism, Yang-Mills and GR) with many canonical momenta serving as auxiliary fields---fields that can be eliminated algebraically using their own equations of motion. Hence there is little need or room for distinctively Hamiltonian ideas.  

One would then expect Lagrangian-equivalent gauge transformations to appear somehow in the constrained Hamiltonian formalism; this expectation is fulfilled using the Rosenfeld-Anderson-Bergmann-Castellani gauge generator \cite{RosenfeldQG,AndersonBergmann,CastellaniGaugeGenerator,SalisburySundermeyerRosenfeldQG}, which combines primary and secondary first-class constraints in a special way.  This gauge generator $G$ preserves Hamilton's equations and changes the canonical action at most by a boundary term.  $G$ generates transformations basically equivalent to the usual 4-dimensional Lagrangian expressions, such as a $4$-gradient in electromagnetism or a space-time coordinate transformation (on-shell) in General Relativity.  Hence the Lagrangian-equivalent Hamiltonian formalism forms a coherent package reflecting a remark by Pons and Shepley regarding a special case: ``We have been guided by the principle that the Lagrangian and Hamiltonian formalisms should be equivalent \ldots in coming to the conclusion that they in fact are.''  \cite{PonsReduce} 

While the acceptability of the gauge transformations generated by $G$ is not in doubt, perhaps it is too restrictive?  Are gauge transformations generated only by $G$, or does each first-class constraint \emph{by itself} generate a gauge transformation? The latter view arose as early as Bergmann \& Schiller \cite{BergmannSchiller,BergmannObservables}, was held by Dirac \cite{DiracLQM}, and probably remains the majority view \cite{HenneauxTeitelboim}, though the stricter Lagrangian-equivalent view has been reviving from the 1980s and is gaining attention  \cite[p. 104]{Kiefer3rd}.  There is, however, widespread agreement that second-class constraints do not generate gauge transformations.

One might think that electromagnetism is too simple and well known to teach us anything about general formalisms for gauge theories at this late date; on the other hand, the fact that electromagnetism is well understood opens the door to using examples to test (not merely illustrate) the formalism, a practice also found in Sundermeyer's book \cite{Sundermeyer}. Moreover, comparing massless (Maxwell) and massive (Proca) electromagnetisms is standard practice in particle physics, but massive theories (whether electromagnetic or gravitational) are traditionally not considered as much among general relativists. Hence this example provides an opportunity to bring general principles and accepted examples further into alignment.  

It might be worthwhile to recall that the separate first-class constraints view implies that observables are inequivalent \cite{ObservablesEquivalentCQG} between massive electromagnetism attributed to Proca $$ \mathcal{L} = - \frac{1}{4} F_{\mu\nu} F^{\mu\nu} - \frac{1}{2} m^2 A_{\mu} A^{\mu}$$ and the gauge version attributed to Stueckelberg $$  \mathcal{L} = - \frac{1}{4} F_{\mu\nu} F^{\mu\nu} - \frac{1}{2} m^2 (A_{\mu} + \partial_{\mu} \phi) ( A^{\mu} + \partial^{\mu} \phi ).$$ 
For Proca, there are no first-class constraints, so everything should be observable including $A_{\mu}.$  The obvious Stueckelberg equivalent is $(A_{\mu} + \partial_{\mu} \phi) $, which therefore should be observable.  Unfortunately $(A_{\mu} + \partial_{\mu} \phi) $ does not have $0$ Poisson bracket with each first-class constraint.  It does, however, have $0$ Poisson bracket with $G$ (making use of the Anderson-Bergmann velocity Poisson bracket \cite{AndersonBergmann}).  That Proca and Stueckelberg massive electromagnetisms have different observables, as the separate first-class constraints doctrine implies,  is difficult to accept considering that these are generally considered to be formulations of the same theory, that  one can gauge-fix the latter into the former using the gauge condition $\phi=0,$ and that the equivalence of the two formulations is used to show that massive electromagnetism is both unitary  and renormalizable  \cite[pp. 738, 739]{PeskinSchroeder}\cite[chapter 21]{WeinbergQFT2}\cite[chapter 10]{Kaku}. The team view of constraints using $G$, on the other hand, leads to equivalent observables between Proca and Stueckelberg. Evidently electromagnetism(s) can still teach us something.

Regarding Maxwell's theory, various conclusions were drawn by Pitts (drawing on the 1980s+ Lagrangian-equivalent reforming literature) about how a first-class constraint by itself does not generate a gauge transformation in any interesting sense \cite{FirstClassNotGaugeEM}. That paper has elicited a detailed response from Pooley and Wallace aiming to vindicate the common view that a first-class constraint generates a gauge transformation  \cite{PooleyWallacePittsGauge,GomesButterfieldMaxwellGauge}. The interested reader can refer elsewhere for a response \cite{PooleyWallaceGomesButterfield}, which includes a concession regarding section 9.


A point of  general interest, however, emerges in response to Pooley and Wallace's calculations, once one generalizes them to permit a photon mass.  That general point, already foreshadowed to some degree earlier \cite[p. 403]{FirstClassNotGaugeEM}, is that \emph{a second-class primary constraint in Proca's massive electromagnetism generates a gauge transformation in the same sense as the first-class primary constraint does in Maxwell's theory}.  It is generally accepted that a second-class constraint (or even a team of them) does not generate a gauge transformation. If there is no good reason to say that the primary constraint generates a gauge transformation in Maxwell's theory but not in Proca's theory, the common view faces a paradox.   Likewise, as will appear below in detail, the second-class primary constraints in massive spin $2$ relatives of General Relativity generate just as much of a gauge transformation as do the first-class primary constraints in GR.  Given that it is generally accepted that second-class constraints do not generate gauge transformations, these results pose a paradox for  the separate first-class constraints not faced by the view involving  the gauge generator $G$, a tuned sum of first-class constraints.  On this Lagrangian-equivalent view, neither a first-class primary constraint in Maxwell or GR nor a second-class primary constraint in Proca or massive gravity generates a gauge transformation, but in the massless theories the primary constraints acquire suitable first-class secondary constraint teammates to make a gauge transformation leaving the action quasi-invariant, whereas in the massive theories the second-class secondary constraints do not make suitable teammates.


\section{De-Ockhamization and Gauge Freedom?} 

The inverse Legendre transformation from the extended Hamiltonian to a Lagrangian with a redefined electrostatic potential \cite{PooleyWallacePittsGauge} is offered by Pooley and Wallace as vindicating the longstanding ``orthodox'' (as they put it) claim that a first-class constraint generates a gauge transformation \emph{individually} in electromagnetism---a  claim disputed previously especially by Pons and collaborators (\emph{e.g.}, \cite{PonsDirac,PonsSalisburySundermeyerFolklore}) and more recently by various arguments by Pitts  \cite
{GRChangeNoKilling,FirstClassNotGaugeEM,ObservablesEquivalentCQG,ObservablesLSEFoP,BergmannObservables}. (Various authors from the 1980s onward have sought to make the Hamiltonian formalism equivalent to the Lagrangian, but much  of this work was not very elenctic.)  
 The fact that one arrives at a de-Ockhamized Lagrangian (splitting one quantity into the sum of two for no evident reason) might give one pause, however.  Has one really learned something about electromagnetism and its gauge freedom?  Or has one merely been treated to an over-interpreted instance of a trick, de-Ockhamization (splitting one quantity into two), that one could apply to any dynamical quantity in any theory, one with first-class constraints, one with second-class constraints, one with both, or perhaps even one with no constraints?  Is an unmotivated de-Ockhamization an interesting form of gauge freedom, or an \emph{ad hoc} maneuver akin to splitting force into ``the sum of gorce and morce'' \cite{GlymourEpist}?

 If a first-class constraint generates a gauge transformation only in non-standard senses such as de-Ockhamization or a canonical transformation  that works just as well for a second-class primary constraint as for a first-class primary constraint \cite[p. 403]{FirstClassNotGaugeEM}, then the late 1950s+ conventional wisdom about separate first-class constraints is not yet vindicated, \emph{pace} Pooley and Wallace.  Neither de-Ockhamization nor canonical transformations require first-class constraints.
 There exist interesting cases of artificial gauge freedom \cite{Arkani,Ruegg,PittsArtificial,FrancoisArtificialSubstantialGauge}, but additive algebraic de-Ockhamization seems not to appear  among them.


\section{Do Second-Class Primary Constraints Also Generate Gauge Transformations? }

Pooley and Wallace appear  not to distinguish  between two quite different claims:  (1) that a first-class constraint really does generate a gauge transformation in electromagnetism, as a \emph{fact} about Maxwell's theory (among others) in Hamiltonian form that (presumably) Dirac and/or Bergmann \emph{discovered}, and (2) that one can devise a noncontradictory  formalism  according to which one can preserve the traditional claim that a first-class constraint generates a gauge transformation as a matter of theoretical choice.  Unfortunately Pitts \cite{FirstClassNotGaugeEM} denied claim (2) in section 9 using a largely verbal rather than mathematical argument, while more or less affirming it as a triviality in section 10 among other places.  Pooley and Wallace aptly argue for claim (2), especially  by performing an inverse Legendre transformation to what one might call an `extended' Lagrangian, a Lagrangian with a de-Ockhamized electrostatic scalar potential.  They do little to support claim (1), however. Neither do they recognize the equivocation between the two.  Kukla has described the process of conflating trivial and nontrivial versions of a claim as a ``switcheroo'' and/or ``reverse switcheroo'' \cite[p. x]{KuklaConstructivism}.

After arriving at the extended Hamiltonian (their equation 27), Pooley and Wallace calculate the resulting Hamilton's equations, which are a modification of the Maxwell equations with the gradient of an arbitrary function added in the case of equation 28 for $\dot{\vec{A}}.$  They immediately add that
\begin{quote}  
[w]e can swiftly verify that the transformations generated by $O_t[f,g]$ [their expression that takes arbitrary combinations of the {separately} smeared primary and secondary constraints] are symmetries of \emph{these} equations for arbitrary $f,$ $g.$
 \end{quote}
But this is too quick. Proceeding more slowly, one sees that one can find symmetries  if and only if one is willing to redefine arbitrary functions in the equations---something that one does not have to do with the uncontroversial Lagrangian-equivalent gauge transformations that transform the $4$-vector potential by a $4$-dimensional gradient.  Hence the extended Hamiltonian requires a lowering of standards in order to call the resulting transformations gauge transformations.  Perhaps one can live with this, as long as these lowered standards do not commit one to finding gauge freedom in places that by common consent do not involve gauge freedom.  Unfortunately for the ``orthodoxy'' that Pooley and Wallace aim to preserve, just such a result occurs, as will now be shown using Proca's massive electromagnetism.  

One gets Proca's theory of massive electromagnetism---commonly described as involving massive photons, though the quantum words are used even for the classical theory---by adding a ``photon mass'' term $- \frac{1}{2} m^2 A_{\mu} A^{\mu} = - \frac{1}{2} m^2 (\vec{A}^2 - V^2)$  \cite[pp. 597-601]{Jackson} \cite[pp. 183-186]{Sundermeyer}. Massive electromagnetism approaches massless (Maxwell) as $m \rightarrow 0,$ both classically and in quantum field theory   \cite{BelinfanteProca,Glauber,BoulwareYM,SlavnovFaddeev,GoldhaberNieto2009,UnderdeterminationPhoton}. 
Using $c=1$ and $\hbar=1$ one can convert a mass into an inverse length (tied to the Compton wavelength).  
 Taking the Legendre transformation, one flips the sign of the mass term, so the Hamiltonian (whether canonical or total/primary) sprouts a term $ \frac{1}{2} m^2  (\vec{A}^2 - V^2)$.
The canonical momentum $\pi_0$ is unaffected, so the primary constraint $C_0 = \pi_0$ is just as in Maxwell's theory.  The dynamics shows that dynamically preserving the primary constraint yields a modified Gauss-like law 
$$ C_1^m = \nabla \cdot \vec{\pi} - \rho + m^2 V.$$
One therefore has the total or primary Hamiltonian (corresponding to equations (9), (11) and (12) of Pooley and Wallace) after dropping a boundary term along the way:
\begin{eqnarray} 
H^m =  \int d^3 x \left[ \frac{1}{2} (\vec{\pi}^2 + \vec{B}^2) +  \lambda \pi_0 + \vec{\pi} \cdot \nabla V + (\rho V + \vec{A} \cdot \vec{J})  + \frac{1}{2} m^2(\vec{A}^2 - V^2) \right] = \nonumber \\
= \int d^3x \left[ \frac{1}{2} (\vec{\pi}^2 + \vec{B}^2) +  \lambda C_0 - V \nabla \cdot \vec{\pi}  + (\rho V + \vec{A} \cdot \vec{J})  + \frac{1}{2} m^2(\vec{A}^2 - V^2) \right] = \nonumber \\
\int d^3x \left[ \frac{1}{2} (\vec{\pi}^2 + \vec{B}^2) +  \lambda C_0 - V ( \nabla \cdot \vec{\pi}  -\rho  + m^2 V)    + \vec{A} \cdot \vec{J}  + \frac{1}{2} m^2(\vec{A}^2 + V^2) \right] \nonumber \\
= \int d^3x \left[ \frac{1}{2} (\vec{\pi}^2 + \vec{B}^2) +  \lambda C_0 - V C_1^m    + \vec{A} \cdot \vec{J}  + \frac{1}{2} m^2(\vec{A}^2 + V^2) \right]
 \end{eqnarray}  
where 
$$ C_1^m =  \nabla \cdot \vec{\pi}  -\rho  + m^2 V = 0$$ 
is the modified massive Gauss law in phase space.  Note that the primary constraint $C_0= \pi_0$ is exactly as in Maxwell's electromagnetism, a point that will prove crucial later.  

This massive Hamiltonian, provided for reference, will not be the starting point for the de-Ockhamized massive Hamiltonian to be found later, because it is not clear what, if anything, one can  do to a Hamiltonian with second-class constraints in order to `find' some kind of gauge freedom.  Indeed one would expect the answer to be ``nothing'' because it is generally agreed that second-class constraints do not generate gauge transformations.  At some point there was a discussion of whether second-class constraints might contribute to the gauge generator with coefficients dependent on the gauge parameters, alongside the first-class constraints; the answer is negative  \cite{ChitaiaGaugeFirstSecond}.  One thing that can happen is that one needs to redefine the constraints, mixing in some terms that naively look like second-class constraints into a modified set of first-class constraints.

However, once one sees how the Pooley-Wallace inverse Legendre transformation from the extended Maxwell Hamiltonian works to achieve an `extended' Lagrangian---one with a de-Ockhamized electrostatic potential $V$---it becomes clear how to make a massive generalization.  One can simply add a de-Ockhamized mass term to the Pooley-Wallace de-Ockhamized massless Lagrangian and then take the Legendre transformation to get the de-Ockhamized (`extended'?) massive Hamiltonian.  
Their de-Ockhamized  Lagrangian density (equation 48) is, after discarding an inessential spatial integration, 
$$ \mathcal{L}_{\mu^{\prime} }[\vec{A}, V; \dot{\vec{A}}, \cancel{ \dot{V}} ] =     \frac{1}{2} (\dot{\vec{A}} - \nabla(V + \mu^{\prime}) )^2     - \frac{1}{2}(\nabla \times \vec{A})^2 - ((V + \mu^{\prime}) \rho + \vec{A} \cdot \vec{J} ). 
$$
This is just the usual Maxwell Lagrangian density, but with $V + \mu^{\prime}$ playing the role of electrostatic potential.  To add a mass term one would usually add $- \frac{1}{2} m^2 ( -V^2 + \vec{A}^2 )$ to the Lagrangian density, but now we need to use the de-Ockhamized electrostatic potential, so the mass term is now 
$$ - \frac{1}{2} m^2 ( - (V + \mu^{\prime})^2 + \vec{A}^2).$$ 
Using the Pooley-Wallace expression plus the de-Ockhamized mass term, one can now perform the Legendre transformation in the usual way as far as possible (recalling that gauge freedom arises precisely because one cannot perform the full Legendre transformation). 
One gets for the canonical momenta
\begin{equation}  
 \vec{\pi}_E = \dot{\vec{A}} - \nabla (V + \mu^{\prime}),
\end{equation} 
which is equivalent to their equation 28, and 
$$\pi_0=0.$$
Calculating the total or primary Hamiltonian density (the one equivalent to the Lagrangian density, in this case equivalent to the de-Ockhamized Lagrangian density) yields, dropping a spatial divergence along the way,
\begin{eqnarray} 
\mathcal{H}_E^m = \frac{1}{2} \vec{\pi}^2 + \vec{\pi} \cdot \nabla (V + \mu^{\prime}) + \frac{1}{2} \vec{B}^2  + \frac{1}{2} m^2 \vec{A}^2  - \frac{1}{2} m^2 (V + \mu^{\prime})^2 + (V + \mu^{\prime}) \rho + \vec{A} \cdot \vec{J} + \lambda \pi_0  = \nonumber \\
\frac{1}{2} \vec{\pi}^2   + \frac{1}{2} \vec{B}^2     - (V + \mu^{\prime}) \nabla \cdot \vec{\pi}     -  m^2 (V + \mu^{\prime})^2 + \frac{1}{2} m^2 (V + \mu^{\prime})^2 + (V + \mu^{\prime}) \rho + \vec{A} \cdot \vec{J}+ \frac{1}{2} m^2 \vec{A}^2 + \lambda \pi_0  = \nonumber \\ 
\frac{1}{2} \vec{\pi}^2   + \frac{1}{2} \vec{B}^2     - [V + \mu^{\prime}) (\nabla \cdot \vec{\pi}    +  m^2 (V + \mu^{\prime})  -  \rho] + \vec{A} \cdot \vec{J}+  \frac{1}{2} m^2 (V + \mu^{\prime})^2 + \frac{1}{2} m^2 \vec{A}^2 + \lambda \pi_0.  
\end{eqnarray} 
For convenience, terms that turn out to be the secondary constraint have been gathered into square brackets.
This expression reduces to the usual Proca Hamiltonian given above for $\mu^{\prime}=0,$ as expected.  Perhaps less obvious is the fact that the de-Ockhamized Hamiltonian obtains simply from de-Ockhamizing the electrostatic potential; no terms involving the momenta are affected.  
While one does not normally talk of an extended Hamiltonian for a theory with no first-class constraints, such as Proca's massive electromagnetic theory, that is merely a verbal issue.  The key feature of the `extended'  electromagnetic Lagrangian (Pooley and Wallace equation 48) is the de-Ockhamization of the electrostatic scalar potential---a process that one can perfectly well apply to the Proca Lagrangian.  
Requiring consistency over time by preservation of the primary constraint $\pi_0=0$ gives the secondary constraint
$$ \nabla \cdot \vec{\pi} + m^2(V + \mu^{\prime}) - \rho,$$ 
so the secondary constraint is de-Ockhamized \emph{via} the expression
 $$(V + \mu^{\prime}).$$
Preserving the de-Ockhamized secondary constraint with the de-Ockhamized Hamiltonian's time evolution gives
$$ - \nabla \cdot \vec{A}  + m^2 \lambda = 0,$$
which is the de-Ockhamized Lorentz condition, as one would expect  \cite[pp. 597-601]{Jackson}---a law, not just a good idea as in the massless case.  Thus no tertiary constraint arises.  
The primary constraint has zero Poisson bracket with itself.  The secondary constraint also has zero Poisson bracket with itself (not always a trivial result in field theory, because one is in fact dealing with infinitely many constraints, one per point, in contrast to particle theories, in which the antisymmetry of the Poisson bracket guarantees that result).  
But the cross-term is not zero: 
$$ \{ \pi_0(\vec{x}), \nabla \cdot \vec{\pi}(\vec{y}) + m^2 (V + \mu^{\prime})(\vec{y}) - \rho (\vec{y}) \}= - m^2 \delta(\vec{x}, \vec{y}).$$
Hence the constraints are second-class due to this cross-term.  This result is, however, a relation between the constraints, not a property of either one of them, a fact that will facilitate doing something unusual with the second-class primary constraint in massive electromagnetism.

The question of the sense in which the transformations generated by separately smeared constraints preserves the Hamiltonian field equations requires more scrutiny.  A standard notion of gauge symmetry is one that preserves the Lagrangian density up to a total divergence (and thus the action up to a boundary term), and thus preserves the Euler-Lagrange equations.  In a Hamiltonian formulation, Hamilton's equations are the Euler-Lagrange equations for the canonical Hamiltonian $\int d^3x (p \dot{q} - H),$ so preserving the Euler-Lagrange equations is preserving Hamilton's equations.  One can check by direct calculation whether a separate first-class constraint preserves the canonical action up to a boundary term in electromagnetism; the answer is negative \cite{FirstClassNotGaugeEM}.  But perhaps there is some interesting weaker notion available, perhaps one formulated in terms of the equations of motion? 

 After presenting the equations for the extended Hamiltonian for Maxwell's (massless photon) electromagnetism, Pooley and Wallace tell us, let us recall, that ``[w]e can swiftly verify that the transformations generated by $O_t[f,g]$ are symmetries of \emph{these} equations for arbitrary $f$, $g$.''  While one actually cannot verify that swiftly (because it does not straightforwardly hold), the effort to do some of it slowly will be  rewarding.  Generalizing their equations (28-31) to admit the photon mass term, noting that simply setting $m=0$ recovers the massless case, one obtains: 
\begin{eqnarray} 
\dot{\vec{A}} = \frac{\delta H_E^m}{\delta \vec{\pi} } = \vec{\pi} + \nabla(V + \mu^{\prime}), \\
\dot{\vec{\pi}} = - \frac{\delta H_E^m}{\delta \vec{A} } =  - \nabla \times \vec{B} - \vec{J} - m^2 \vec{A}, \\ 
\dot{V} = \frac{\delta H_E^m}{\delta \pi_0} = \lambda, \\ 
\dot{\pi_0} = - \frac{\delta H_E^m}{\delta V} = \nabla \cdot \vec{\pi} - \rho + m^2(V + \mu^{\prime})=0.
\end{eqnarray} 
The primary constraint leaves the canonical momenta and $\vec{A}$ alone and changes the electrostatic potential $V$ by the arbitrary smearing function for both $m=0$ and $m \neq 0.$  The secondary constraint sprouts a new term for $m\neq 0,$ altering the canonical momentum $\pi_0$, which is bad, because that canonical momentum should stay $0$.  But one never expected anything gauge-like to arise from the second-class secondary constraint in massive electromagnetism, so this spoilage is  the kind of thing that one expected  \cite[pp. 183-186]{Sundermeyer}.  The effect of the primary constraint, namely shifting $V$ by the arbitrary smearing function $f$, leads to the following modified Hamiltonian field equations: 
\begin{eqnarray} 
\dot{\vec{A}}  =  \vec{\pi} + \nabla(V + \mu^{\prime} + f), \\
\dot{\vec{\pi}} =  - \nabla \times \vec{B} - \vec{J} - m^2 \vec{A}, \\ 
\dot{V} + \dot{f}  =  \lambda, \\ 
\dot{\pi_0} =  \nabla \cdot \vec{\pi} - \rho + m^2(V + \mu^{\prime} + f )=0.
\end{eqnarray} 
The first equation shows that the extended Hamiltonian sets a new, lower standard of having to further redefine the electrostatic potential by $f,$ rather than leaving the equation invariant as the standard Lagrangian-equivalent gauge transformations do.  Perhaps one can live with this lowered standard, as long as one implements it consistently.  The second equation is invariant.  The third equation involves redefining the arbitrary function $\lambda,$ which is not surprising given that this arbitrary function corresponds to the velocity of the electrostatic potential.  The fourth equation, however, contains a disappointment for the conventional (separate first-class constraints) view, because it shows that the \emph{ massive ($m\neq0$) equations are just as invariant under the transformation  generated by the primary constraint $\pi_0$ as the massless ($m = 0$) equations}:  one alters the de-Ockhamized electrostatic potential in the same way in equation 4 in the mass term as one does in equation 1 shared with the massless case.

  For proponents of the extended Hamiltonian, the primary first-class constraint generates a gauge transformation for Maxwell's theory, suitably formulated with a de-Ockhamized electrostatic potential, but no second-class constraint is expected to generate a gauge transformation.  
 Likely no one accepts such a result---neither the mathematically strict Hamiltonian-Lagrangian equivalent view  \cite{RosenfeldQG,AndersonBergmann,CastellaniGaugeGenerator,PonsDirac,SalisburySundermeyerRosenfeldQG,FirstClassNotGaugeEM} nor the conventional separate first-class constraints view \cite[pp. 19-21]{DiracLQM} holds that a second-class constraint generates a gauge transformation. One might take this to pose a paradox for the idea that de-Ockhamizing the electrostatic potential in electromagnetism introduces nontrivial gauge freedom.  In any case the conventional view that a first-class constraint (by itself) generates a gauge transformation (apart from a class of exceptions of no relevance here  \cite[p. 19]{HenneauxTeitelboim}) \emph{and a second-class constraint does not generate a gauge transformation}, faces a puzzle.  

The procedure employed to find (install?) gauge freedom in Proca's electromagnetism \emph{via} de-Ockhamization result differs in at least two key ways from Dirac's process of  extending the Hamiltonian for Maxwell's theory.  First,  only the primary (second-class) constraint, not the secondary (second-class) constraint, generates a gauge transformation (by the standards at hand).  Instead of Pooley and Wallace's $O[f,g] = \int (f C_0 - g C_1)$ that separately smears the primary and secondary constraints with arbitrary functions, I use an arbitrary smearing function $f$ for the primary, but set $g=0,$ omitting the secondary constraint from the game. The argument does not need to show that every second-class constraint generates a gauge transformation by the Pooley-Wallace criterion, but only that one of them does.  That is a good thing, because in fact only one of them, the primary constraint, actually does so.  Second, one does not simply add terms in the secondary constraints to the Hamiltonian as in Dirac's extension procedure.  Instead one does whatever is required by de-Ockhamizing the electrostatic potential.  For the massless case, de-Ockhamization implies adding the secondary constraint(s) with an arbitrary coefficient(s); for the massive case it does not, because there are quadratic terms in the electrostatic potential to address.  

One might wonder why the second-class character of the primary constraint in Proca's massive electromagnetism does not spoil whatever `extended' gauge freedom might exist from the primary constraint in the massless theory.  The primary constraint is the same in Proca's theory as in Maxwell's because the kinetic term, the piece of the Lagrangian with derivatives, is the same.  The primary constraint generates exactly the same transformation, adding an arbitrary smearing function to the electrostatic potential, whether or not there is a photon mass term.  Crucially, considered in \emph{isolation} there is nothing second-class about the primary constraint $\pi_0$:  its Poisson bracket with itself vanishes.  It is called second-class because of a \emph{relation} to the massive Gauss secondary  constraint.  But the massive Gauss constraint plays no role in this part of the de-Ockhamization process. Consequently it simply does not matter whether the primary constraint is first-class or second-class.  Thus this trick manages to generate a gauge transformation from the second-class primary constraint in Proca by the same standard that Pooley and Wallace employ in finding a gauge transformation generated by the first-class primary constraint in Maxwell.  

One can further understand the issue 
as follows.  The uncontroversial gauge transformations of the standard Maxwell Lagrangian density $- \frac{1}{4} F_{\mu\nu} F^{\mu\nu}$ are generated by the gauge generator $G$, a tuned sum of the secondary first-class constraint (phase space Gauss law) to modify $\vec{A}$ by a spatial gradient and the primary first-class constraint $\pi_0$, which changes the electrostatic potential by a time derivative of the same smearing function (up to a sign).  A photon mass term, if introduced, breaks this gauge freedom.  On the other hand, de-Ockhamization of the Maxwell Lagrangian introduces a new gauge freedom of sorts, generated by the primary constraint $\pi_0,$ which changes the electrostatic potential by itself in an arbitrary way.  As it happens, introducing a photon mass term and de-Ockhamizing the electrostatic potential do not get in each other's way.  So one can de-Ockhamize the electrostatic potential in massive electromagnetism and thus introduce a new gauge freedom of sorts, generated by the primary constraint $\pi_0$, which changes the electrostatic potential by itself in an arbitrary way.  The massive theory's second-class secondary constraint (modified phase space Gauss law, including a mass term bit) is not used.  Because $\{ \pi_0(x), \pi_0(y) \}$ = 0, the primary constraint considered \emph{all by itself} (as if the secondary constraint did not exist) is, so to speak, first-class with itself. A primary constraint in electromagnetism is not \emph{per se} first- or second-class, but only in relation to the secondary Gauss law constraint.  Thus for massive electromagnetism, the second-class nature of $\pi_0$ is not a property of $\pi_0,$ but a \emph{relation} with the secondary constraint (the Gauss law), and that offending relatum is not used in a relevant way (as a generator) in the de-Ockhamization process.  Thus one sees why, if one accepts the conventional view that a first-class constraint generates a gauge transformation in Maxwell's electromagnetism, one is also feels pressure  by the same standard to accept that  a second-class primary constraint generates a gauge transformation in Proca's electromagnetism.  On the other hand, for the Lagrangian-equivalent view involving the gauge generator $G$, neither the first-class primary in Maxwell's theory nor the second-class primary in Proca's theory generates a gauge transformation, so one does not feel pressure to accept that a second-class primary constraint generates a gauge transformation.  


\section{De-Ockhamizing the Equations of Motion or the Action?} 

While the Pooley-Wallace calculation works at the level of the equations of motion, one can equally well apply a primary constraint (first-class or second-class) to the action and see that it performs analogously---it de-Ockhamizes $A_0$ or the analogous quantities (such as the lapse function and shift vector in GR).  For Maxwell's electromagnetism, one can find the calculation, for example, in (\cite[p. 394]{FirstClassNotGaugeEM}).  There one sees that the action changes by a term proportional to the phase space Gauss law.  It was not remarked there, but can be seen readily, that in effect one replaces $A_0$ by $A_0$ plus or minus the arbitrary smearing function.  This is not a gauge transformation in the uncontroversial  sense of changing the action by at most a boundary term and hence preserving the Euler-Lagrange equations (Hamilton's equations).   

What happens for Proca's theory, where one ought to find no gauge freedom but, as has just appeared above, finds gauge freedom 
 by the Pooley-Wallace criterion? The Maxwell calculation needs to be supplemented by the effect of the primary constraint on the mass term.  (I use $-+++$ signature.)  
\begin{eqnarray}  \{ \int d^3y \xi(t,y) p(y), - \frac{1}{2} m^2 A_{\mu} A^{\mu} (t,x) \}  = 
 m^2 \xi(t,x)  A^0(x) = -m^2 \xi A_0 .\end{eqnarray}   In the mass term also, one has simply shifted $A_0$ by the arbitrary smearing function; one can get the quadratic term $\frac{m^2}{2} \xi^2$ from the quadratic term in exponentiating the Poisson bracket with the smeared primary constraint, thus accommodating finite gauge transformations.
  Is there a reason that  this de-Ockhamization  counts as a gauge transformation for Maxwell's theory and not in Proca's theory?  Working with the action, we have now had success with a quadratic expression in $A_0$ with a constant coefficient.


\section{Application to General Relativity and Massive Gravities}

One might perhaps wonder whether the point involving massive electromagnetism relies on some respect in which electromagnetism is simple.  Does a similar result obtain in massive gravity? If so, that  bears directly on quantum gravity. Because the Lagrangian-equivalent Hamiltonian formalism includes space-time coordinate transformations (on-shell), the argument also provides support for representing space-time coordinate transformations, not merely spatial ones, in a canonical formalism.  Spatio-temporal covariance  being a key feature of GR, the spirit of GR is better manifest in the Lagrangian-equivalent Hamiltonian formalism using $G$ as opposed to separate first-class constraints as gauge generators.

It is of interest, therefore, not to try to  rely on whatever analogies  exist between electromagnetism and gravity, but rather to ascertain by direct calculation whether the result about second-class primary constraints in massive electromagnetism also works for massive spin $2$ gravities, relatives of GR.  The Pooley-Wallace-esque argument would go like this:  in GR the Hamiltonian field equations are linear (in the broad sense of affine) in the Lagrange multipliers $N$ and $\beta^i$ (the ADM lapse and shift \cite[chapter 21]{MTW}), so applying the first-class primary constraint (with an arbitrary smearing function\footnote{Smearing functions here are often written as functions of $x$ or $y$ or $z$.  That does not exclude dependence on time $t$, which generally is present.  The expressed dependence on $x$ or $y$ or $z$ is responsive to the demands of the functional mathematics used  \cite{Sundermeyer}.  Due to the  merely algebraic dependence on the lapse and shift in the calculations done here, the functional mathematics reduces to partial differentiation.}) to the GR Hamiltonian field equations shifts $N$ and $\beta^i$ in an arbitrary way given by the smearing function, which constitutes a gauge transformation.  

As it happens, one can readily enough find a massive relative or two of GR (retaining Lorentz covariance) that, like Proca's electromagnetism, has an action quadratic in the $N$ and/or $\beta^i$ (which are now auxiliary fields rather than Lagrange multipliers). That requirement is merely for illustration, because the same points apply for expressions with arbitrary algebraic dependence on the lapse and shift, as will appear below.

An interesting family of massive gravity theories, mostly containing spin $0$ ghosts (but some of which were reinvented around 2010  and shown to be nonlinearly pure spin $2$ \cite{deRhamGabadadze,HassanRosenNonlinear}), is the 2-parameter family due to Ogievetsky and Polubarinov \cite{OP} in 1965.\footnote{This paper has the striking quality of making important conceptual innovations in geometry, wholly using  the language of perturbative series expansions.  Given the well-known familiarity  of general relativists with geometry and of particle physicists with  perturbative series expansions \cite{BrinkDeserSupergravity}, the paper could perhaps build bridges between communities, but also could prove difficult for either side to grasp fully.  Some notational choices could perhaps have made the paper more accessible. A recent paper aims to introduce the OP paper to modern audiences \cite{OPMassiveGravityBenign}. }  
  The two parameters are (give or take a sign) the density weight of the potential and the power (even non-integral!) to which the densitized metric is raised. Those authors impose a ``spin limitation principle'' to exclude spin $1$ and one spin $0$ from the field equations.  Their Poincar\'{e}-invariant (or sometimes more) mass terms take the form
\begin{eqnarray} 
\frac{m^2}{32 \pi G n}  \left( -2(1 - 2 l) \sqrt{-g}   - \frac{1}{n} (\sqrt{-g}^l g^{\mu\nu} \eta_{\nu\alpha})^n \delta^{\alpha}_{\mu} + \frac{2}{n}[(1-2l)n + 2] \right),
\end{eqnarray}
where I have taken the liberty of replacing their $x^4 = ict$ with real coordinates and the Minkowski matrix $\eta_{\mu\nu} = diag(-1,1,1,1).$
This expression is to be understood using the binomial series expansion to make sense of non-whole powers of   $ (\sqrt{-g}^l g^{\mu\nu} \eta_{\nu\alpha})^n.$ 
(The power $n$ is therefore of the inverse metric density.  One should also note a roughly East-West difference over the definition of density weight; I have used $l = -p$ to replace their parameter $p$ with one matching the bulk of Western texts.) 
For $n=1$ (contravariant with weight $l$), the field is the weight $l$-densitized contravariant/inverse metric; for $n=-1$ and parameter $l$, the field is $\eta^{\mu\alpha} \sqrt{-g}^{\, -l} g_{\alpha\beta} \eta^{\beta\nu}$.  Other powers of $n$ will give polynomials or (typically) infinite series.  It will not be necessary to study $n \neq \pm 1$ theories explicitly here, but they and even more general theories will be included in a more abstract derivation below.  The Ogievetsky-Polubarinov work (see also \cite{OP}) is probably the first published use of nonlinear group realizations, as well as the first clear statement of how spinors can be spinorial with respect to (nearly) general coordinate transformations, obviating an orthonormal tetrad \cite{OPspinor,PittsSpinor}. One can find similar ideas in the 1949 dissertation of Bryce DeWitt \cite{DeWittDissertation,DeWittSpinor1950}.

The mass term above is physically normalized.  It has become customary to write Hamiltonian GR using a geometric rather than physical normalization \cite[p. 520]{MTW}, yielding the elegant expression $ N \mathcal{H}_0 + \beta^i \mathcal{H}_i$ 
(along with some terms in the primary constraints  \cite{Sundermeyer} that are often neglected).  One could try restoring the physical normalization to the kinetic term (the GR expression), but it is easier and more familiar to retain the geometric normalization and apply it to the mass term also by multiplying by $16 \pi G$ \cite{DeserMass,MassiveGravity1}.

In these massive gravity theories, the cosmological constant gives zeroth, first and higher order terms in the action (hence nothing, zeroth and first and higher order in  the field equations), while tuned linear and constant terms in the action cancel them out the zeroth and first order terms in the action (hence the zeroth order term in the field equations, which gives the peculiar growing-with-distance behavior of the cosmological constant \cite{FMS,LambdaMPIWG}).  What remains is a linear term (and higher) in the field equations, with exponential Yukawa falloff.  Obviously the graviton mass must be very small indeed if such theories are to have any hope of fitting the actual world.\footnote{Various devils in the details were discovered in the early 1970s   \cite{vDVmass1,vDVmass2,DeserMass}.  To some degree they were resolved at the time  \cite{Vainshtein,Vainshtein2,MaheshwariIdentity,MassiveGravity3}, but with little immediate influence.  Hence research was nearly dormant for decades outside the school of A. A. Logunov (\emph{e.g.}, \cite{LogunovBasic}), which advocated the spin $2$-spin $0$ Freund-Maheshwari-Schonberg theory \cite{FMS} and underestimated the seriousness of the spin $0$ ghost problem  under quantization. A  curious feature is that the Tyutin-Fradkin-Boulware-Deser nonlinear spin $0$ ghost  \cite{TyutinMass,DeserMass} was pre-eliminated for one massive theory by  Masheshwari (a student of Freund at Chicago) \cite{MaheshwariIdentity,MassiveGravity3} before it was discovered as supposedly generic and, 40 years later, re-eliminated for a broader collection  massive Poincar\'{e}-invariant theories.  This is a striking  case of the contingency of the trajectory of fundamental gravitational/high-energy physics. Maheshwari's result was not noticed partly because he  changed his research interests to a slower-paced topic more suited to his teaching duties and lack of journals and preprints in a new regional Indian university \cite{MaheshwariBiography}, that topic being functional integration. When massive gravity should have been regarded as a viable project, or even whether it ever should have been, is not relevant for the main task of this paper, however. It is not required that massive gravity theories avoid ghosts, tachyons, discontinuous massless limits, or any other indignities that might exist, given the uses at hand.}  Actually one might rather speak of graviton mass\emph{es}, because the theories permit the spin $0$ (with negative energy) to be lighter or heavier than the spin $2$.  \emph{If} one could somehow tame the spin $0$ ghost, possibly its distinct mass could be phenomenologically useful \cite{GrishchukMass}.  But such considerations are not important here, where it is important only to have a Lagrangian that differs from GR by some algebraic terms with certain properties, not that the theory admit stable quantization.   That later situation might not be so obviously doomed, however \cite{ParkMassiveGravity2,OPMassiveGravityBenign}. 

One can now introduce the ADM space-time split into space and time, expressing the (inverse) metric and determinant in terms of the spatial metric $h_{ij} = g_{ij}$, lapse $N$ and shift $\beta^i$ as follows.  In an ADM split \cite{MTW} (pp. 506-8), the lapse function $N(x,t)$ describes the relation between physical time and coordinate time.  For a world-line orthogonal  to a given simultaneity slice, a unit temporal coordinate length  corresponds to a proper physical time lapse of $N,$ justifying the name ``lapse function'' for $N.$  The shift vector field $\beta^{i}(x,t)$ describes how the spatial coordinate system shifts relative to the time-like vector orthogonal  to the simultaneity hypersurfaces. (For taking Poisson brackets, it is important to make a consistent choice of primitive variable $\beta^i$ or $\beta_i$ (indices being moved by the spatial metric) or maybe something more exotic, a choice that might depend on the graviton mass term for a massive gravity theory.  It is typically advisable in my view to prefer the contravariant version $\beta^i$ over $\beta_i$ on conceptual and technical grounds, but the two criteria can come apart, as will appear below.)  The spatial metric $h_{ij}$ is merely the spatial part of the space-time metric $g_{\mu\nu}$. The spatial indices are lowered by $h_{ij} $ and raised by the inverse spatial metric $h^{ij}.$ The space-time metric $g_{\mu\nu}$ and inverse space-time metric $g^{\mu\nu}$ and inverse space-time metric are given by, respectively,

\begin{eqnarray}
\left[
\begin{array}{cc}
g_{00} & g_{0j} \\ 
g_{i0} & g_{ij} 
\end{array}
\right]
=
\left[
\begin{array}{cc}
- N^{2} + \beta^{m} \beta^{n} h_{mn}   & \beta^{k} h_{kj} \\
 \beta_{k} h_{ik} & h_{ij} 
\end{array}
\right],    &
\left[
\begin{array}{cc}
  g^{00} & g^{0j} \\ 
g^{i0} & g^{ij} 
\end{array}
\right]
=
\left[
\begin{array}{cc}
 -N^{-2} & \beta^{j}N^{-2} \\ 
 \beta^{i}N^{-2} & h^{ij} - \beta^{i}\beta^{j}N^{-2}
\end{array}
\right]
\end{eqnarray}
Finally, the determinant $g$ of the components of the space-time metric (a space-time scalar density of weight $2$) is related to the lapse and the determinant $h$ of the spatial metric by 
$$\sqrt{-g} = N\sqrt{h}.$$ The lapse is taken to be positive and is of order $1$, like $g_{\mu\nu}$ and $h_{ij}$ but unlike the shift $\beta^j$, which is small when the coordinates and the physics are `weak'.   Any pair of repeated indices is summed over, from $0$ to $3$ for Greek indices or $1$ to $3$ for Latin indices. In terms of the behavior of matter, which is coupled to $g_{\mu\nu}$, the lapse and shift have the same significance (as fragments of the curved metric) as in GR, but the mass term destroys general covariance and yields a more or less harmonic/DeDonder  condition as a consequence of the field equations.  


One can now readily expression the $n=1$ (contravariant tensor (density)) and $n=-1$ (covariant tensor (density)) mass terms, fixing one parameter out of the two-parameter family of Ogievetsky-Polubarinov massive gravities, in terms of the ADM split. 
It is convenient to consider the inverse metric with weight $l$ and the metric with  weight $-l$ in order to make them inverses.   
For the weight $-l$  metric density one has
\begin{eqnarray}
\left[
\begin{array}{cc}
 \sqrt{-g}^{-l} g_{00} & \sqrt{-g}^{-l} g_{0j} \\ 
\sqrt{-g}^{-l} g_{i0} & \sqrt{-g}^{-l} g_{ij} 
\end{array}
\right]
= 
\left[
\begin{array}{cc}
 - N^{2-l} \sqrt{h}^{-l} + \beta^{m} \beta^{n} h_{mn} \sqrt{h}^{-l} N^{-l}   & N^{-l} \sqrt{h}^{-l} h_{jk} \beta^{k} \\
 N^{-l} \sqrt{h}^{-l} h_{ik} \beta^{k} & h_{ij} \sqrt{h}^{-l}  N^{-l}
\end{array}
\right]_.    
\end{eqnarray}  
(These powers of the lapse are a good reason to avoid the common notation $N^i$ for the shift vector, at least in massive gravity.)
For the weight $l$ inverse metric one has
\begin{eqnarray}
\left[ 
\begin{array}{cc}
  \sqrt{-g}^l g^{00} & \sqrt{-g}^l g^{0j} \\ 
\sqrt{-g}^l g^{i0} & \sqrt{-g}^l g^{ij} 
\end{array}
\right]
=
\left[
\begin{array}{cc}
 -N^{l-2} \sqrt{h}^l  & \beta^{j}N^{l-2} \sqrt{h}^l \\ 
 \beta^{i}N^{l-2} \sqrt{h}^l  & N^l \sqrt{h}^l h^{ik} - \beta^{i}\beta^{j}N^{l-2} \sqrt{h}^l
\end{array}
\right]_.
\end{eqnarray}
The off-diagonal terms, most conspicuously those linear in the shift, will drop out after taking the trace with the Minkowski matrix $\eta_{\mu\nu} = diag(-1,1,1,1).$

For the covariant metric density of weight $-l$ (with indices raised by $\eta$), one has $$ (\eta^{\alpha\mu} \sqrt{-g}^{-l} g_{\mu\nu} \eta^{\nu\beta}) \eta_{\alpha\beta} = \sqrt{-g}^{-l} g_{\mu\nu} \eta^{\mu\nu} = N^{2-l} \sqrt{h}^{-l}  -\beta^{i} \beta^{j} h_{ij} \sqrt{h}^{-l} N^{-l}   +  h_{ij} \sqrt{h}^{-l}  N^{-l} \delta^{ij}. $$ Here $ \delta^{ij}$ is just the spatial identity matrix.

The expression $N \mathcal{H}_0 + \beta^i \mathcal{H}_i$ from GR imposes some restrictions on what can be a simple (such as quadratic) dependence on  the lapse and shift in a graviton mass term.  Suppose that we want an expression that is quadratic (and perhaps linear and constant) in the lapse $N$ (perhaps with some admixture of $\sqrt{h}$) and/or quadratic (and perhaps linear and constant) in the shift $\beta^i$ (perhaps with some admixture of $h_{ij}$ but nothing else such as spatial background geometry $\delta_{ij}$ or the lapse).  Such a restriction, made to facilitate calculations, will restrict the choice of $l$, so that only a certain density weight works for the contravariant family and a certain other density weight works for the covariant family. (If one is willing to shift the terms around in  the quadratic formula, one can find lapse dependence $N^1$, $N^0$, $N^{-1}$ to be soluble using the quadratic formula \cite{FMS,DeserMass,MassiveGravity1}, but one gets an unpleasant square root.)\footnote{While linearity \emph{vs.} nonlinearity makes sufficient sense in massive gravity, one encounters all sorts of interesting nonlinear possibilities \cite{OP,deRhamGabadadze,HassanRosenNonlinear}, as well as comparably reasonably field definitions in terms of $N$ \emph{vs.} $n=N-1$, so one wants to be careful when discussing powers.  One might think that, on dynamical grounds \cite{KhazinShnolBook} or in the spirit of covariant perturbation theory, one should use variables that are small when nothing much is happening.  If so, one would avoid using the metric in favor of the metric perturbation $g_{\mu\nu} - \eta_{\mu\nu}$ or some similar expression, perhaps rescaled appropriately with something like $\sqrt{32 \pi  G}$ \cite{OP}, and correspondingly not $N$ but $n=N-1$ or some rescaling thereof.  For positive powers it does not matter all that much---an expression exactly cubic in $N$ is cubic (and quadratic, linear and constant) in $n$:  $N^3 = (n+1)^3$.  Likewise an expression exactly cubic in $n$ is cubic (and quadratic, linear and constant) in $N$:  $n^3 = (N-1)^3$.  For negative or fractional powers, things change.  The expression $N^{-1}$ might reasonably appear in an action \cite{OP,FMS}, but it gives an infinite series in $n$, whereas $n^{-1}$  would be a strange and dangerous thing to behold in an action, diverging in flat space-time in Cartesian coordinates.   }

In order to exclude powers of $N$ larger than $2$ or smaller than $0$ for the covariant ($n=-1$) theory, one sets $l=0$: only the non-densitized metric $g_{\mu\nu}$ qualifies. Hence this is the Ogievetsky-Polubarinov theory with $n=-1$ and $p=-l=0$.  The full mass term is therefore
$$
\frac{m^2}{32 \pi G} (2 \sqrt{-g} - g_{\mu\nu} \eta^{\mu\nu} +2) =   \frac{m^2}{32 \pi G} ( 2 N \sqrt{h} - N^{2}  + \beta^{i} \beta^{j} h_{ij}  -  h_{ij} \delta^{ij} + 2 ). 
$$

For the contravariant ($n=1$) metric density of weight $l$ one finds the trace $$ \sqrt{-g}^l g^{\mu\nu} \eta_{\mu\nu} =  N^{l-2} \sqrt{h}^l  + N^l \sqrt{h}^l h^{ij} \delta_{ij} - \beta^{i}\beta^{j}N^{l-2} \sqrt{h}^l \delta_{ij}.$$ In order to have powers of the  lapse no less than $0$ and no more than $2$, one sets $l=2.$  
The full mass term is therefore
$$ \frac{m^2}{32 \pi G} ( 6 \sqrt{-g} - (-g)g^{\mu\nu} \eta_{\mu\nu} -2 ) = \frac{m^2}{32 \pi G} ( 6 N\sqrt{h} - h - N^2 {h} h^{ij} \delta_{ij} + \beta^{i}\beta^{i} h -2 ).
$$

These mass terms have been chosen in order to get expressions at most  quadratic in the lapse and/or shift (with \emph{constant} coefficients for the quadratic terms) for convenience.  $g_{\mu\nu}$ is purely quadratic in $N$; if one redefines the shift by mixing in some $h_{ij}$ as  $\beta[i] = \beta_{i} \sqrt{h^{ij}},$ which one could take as primitive, then one gets a purely quadratic expression in the modified shift as well. A purely quadratic expression, that is, one not multiplied by other field variables, avoids quadratic shift terms in the equations of motion. The index on this modified shift vector  $\beta[i] $ is midway between up and down, because its spatial transformation properties are midway between covariant and contravariant---something that makes sense in light of  Ogievetsky and Polubarinov's work \cite{OP} using binomial series expansions. For orthogonal coordinates, this kind of  entity gives the ``physical components'' of vectors that one finds in vector analysis and electromagnetism textbooks  \cite{DavisSnider,Griffiths,Jackson}, which are relative to an orthonormal basis aligned with the coordinate basis vectors.   $\beta[i]$ by itself is not a spatial geometric object, but it forms part of one along with the spatial metric $h_{ij}$ \cite{Tashiro1,TashiroYano,SzybiakLie,PittsSpinor}; that is, knowing $\beta[i]$ in one spatial coordinate system is not enough to infer its components in another spatial coordinate system, because one also needs the components of $h_{ij}.$  That does not seem like much of a problem for most purposes, especially if key resulting  quantities (such as some relative of $\mathcal{H}_i$ along the lines of $\mathcal{H}_i \sqrt{h^{ij} }$) is equated to $0$.  Fixating on quadratic expressions in the lapse and/or shift is largely a matter of expository convenience anyway, as will be shown below using a more abstract notation and more general mass terms.

The other mass term, based around $(-g) g^{\mu\nu} \eta_{\mu\nu}$, is more convenient in terms of the shift vector, but less so in terms of the lapse.  Regarding the lapse this is a peculiar case because $(-g)g^{00} = -h,$  so the time-time component does not yield the expected lapse-like auxiliary field after all.  There is an expression quadratic in the lapse that arises from the \emph{trace} of the curved spatial metric, but that is not so useful.  One can tolerate mixing in $\sqrt{h}$ with $N$ as long as the latter does not disappear. Mixing in more of $h_{ij}$, however, such as taking a trace with the flat background geometry $\delta_{ij},$ is underwhelming;  trying to hide it here through a field redefinition will cause it to resurface in the term $N \mathcal{H}_0,$ so there is little point.  Thus this theory will not  give us  a close parallel with GR regarding the lapse.  However, this mass term is purely quadratic in a weight $1$ densitized shift $\tilde{\beta}^i = \beta^i \sqrt{h}$.  Thus one will see easily that the primary second-class constraints conjugate to the (densitized) shift generate a gauge transformation as much as the primary first-class constraints conjugate to the (densitized) shift do in GR (which is not at all, in my view). It might be useful to see that result without the novel complexity of $\sqrt{h^{ij}}$, so a close parallel to GR with simple enough  mathematics, at least regarding the shift (or something close enough), is worth having.

Thus without working very hard one can find a Poincar\'{e}-invariant massive gravity theory quadratic in the lapse (and quadratic in a shift-like variable if one works harder) and another Poincar\'{e}-invariant massive gravity theory quadratic in the (weight $1$) shift, in which one has a precise parallel to the situation with Proca's massive electromagnetism.  There the photon mass term is quadratic in $A_0$ with a constant coefficient, leading to linear field equations.  
Let us see these parallels to massive electromagnetism in more detail.

What does the primary constraint $p(x)$ do to the canonical action for GR, or indeed for a massive gravity theory?  Thanks to the $3+1$ formalism the calculation is not difficult.  One has the canonical Lagrangian (from which Hamilton's equations follow as Euler-Lagrange equations) 
$$\mathcal{L}_c =  \pi^{ij} \dot{h}_{ij} + p\dot{N} + p\dot{\beta}^i - (N \mathcal{H}_0 + \beta^i \mathcal{H}_i - 16 \pi G \mathcal{L}_{mass} +  p\dot{N} + p_i \dot{\beta}^i) = 
  \pi^{ij} \dot{h}_{ij}  - N \mathcal{H}_0 - \beta^i \mathcal{H}_i + 16 \pi G \mathcal{L}_{mass} , $$ so
$$ 
\delta \mathcal{L}_ = \{ \int d^3x p(x) \xi(x), \mathcal{L}_c(y) \} = - \xi(x) \frac{\partial \mathcal{L}_c }{\partial N}(x).$$ 
At this level of abstraction one already sees that the primary constraint is just the infinitesimal generator for translations of the lapse $N$ by $ - \xi,$ effecting a replacement $N \rightarrow N- \xi$ for $\xi$ small enough to neglect quadratic and higher order terms.  If one wants to recover those higher order terms, one can exponentiate the process of taking the Poisson bracket, as will be discussed in more detail below.  Note that whether $p(x)$ is first-class or second-class plays \emph{no role} in the derivation.  Thus one sees that the primary constraint  involving the momentum $p$ conjugate to $N$ generates a gauge transformation in massive gravity just as much as it does in GR.  
Now $N - \xi$ plays the role of $N$, and one can change $N$ more or less arbitrarily (presumably staying positive) while compensating by making the same change in $ \xi.$   The point might be more lively when worked out more explicitly for some specific mass terms.


\section{Massive Gravity Quadratic in Lapse and the Primary Constraint:  Covariant Weight $0$ Theory} 

The relevant mass term is 
$$
\frac{m^2}{32 \pi G} (2 \sqrt{-g} - g_{\mu\nu} \eta^{\mu\nu} +2) =   \frac{m^2}{32 \pi G} ( 2 N \sqrt{h} - N^{2}  + \beta^{i} \beta^{j} h_{ij}  -  h_{ij} \delta^{ij} + 2 ). 
$$
 The mass term does not affect the process of performing the 60\% Legendre transformation that GR admits \cite{Sundermeyer}, but the Hamiltonian formulation will reverse the sign of the Lagrangian mass term.  Hence one has 
$$ \mathcal{H}_{ms} =  N \mathcal{H}_0 + \beta^i \mathcal{H}_i -\frac{m^2}{2}(2N \sqrt{h} - N^2 + \beta^i \beta^j h_{ij} - h_{ij} \delta^{ij} + 2) + \dot{N}p + \dot{\beta}^i p_i.$$  
The kinetic term being shared with GR, one gets the same primary constraints $p(x)$ conjugate to $N$ and $p_i(x)$ conjugate to $\beta^i$.  One has the usual Poisson brackets $\{ N(x), p(y) \} = \delta(x,y),$ \emph{etc.}  
This Hamiltonian is cleanly quadratic in the lapse $N$, so the constraint comprised of the vanishing of $p$ will be of most interest. 

Requiring the preservation of the constraints by the dynamics yields GR-like secondary constraints but with new mass term contributions:
$$ \{ p(x), \int d^3y \mathcal{H}_{ms}(y) \} = - \frac{\partial \mathcal{H}_{ms} }{\partial N}(x) = -\mathcal{H}_0 + m^2 \sqrt{h} - m^2 N =   0$$ and 
$$ \{ p_i(x), \int d^3y \mathcal{H}_{ms}(y) \} = - \frac{\partial \mathcal{H}_{ms} }{\partial \beta^i}(x) = -\mathcal{H}_i(x) + m^2 h_{ij} \beta^j =   0.$$  
These expressions are linear in the auxiliary fields (not Lagrange multipliers as in GR) $N$ and $\beta^i,$ so the lapse and shift are fixed by the dynamics and there are no tertiary or higher constraints.  The dynamics enforces conditions at least vaguely resembling the harmonic/DeDonder condition.

The canonical Lagrangian is 
$$ \mathcal{L}_{c} =  \pi^{ij} \dot{h}_{ij} - N \mathcal{H}_0 - \beta^i \mathcal{H}_i +\frac{m^2}{2}(2N \sqrt{h} - N^2 + \beta^i \beta^j h_{ij} - h_{ij} \delta^{ij} + 2)  .$$ 
It changes  under the smeared primary constraint by an expression that is not a total divergence, so the result is not a gauge transformation in the strong  uncontroversial sense.  The expression for the change is
$$ \delta \mathcal{L}_c = \{ \int d^3x p(x) \xi(x), \mathcal{L}(y) \} = - \xi(y) ( - \mathcal{H}_0 + m^2 \sqrt{h} - m^2 N)(y) = - \xi \frac{\partial \mathcal{L}_c }{\partial N}(y).$$
Thus $\mathcal{L}_c + \delta \mathcal{L}_c$ is just, up to terms quadratic in $\xi$, the Lagrangian $\mathcal{L}_c (N \rightarrow N - \xi).$  If one wants to get the quadratic term also, one can exponentiate the Poisson bracket-taking.  The next term is $\frac{1}{2} (-\xi)^2 \frac{\partial^2 \mathcal{L}_c }{\partial N^2} = -m^2 \frac{1}{2} \xi^2 ,$ which is just the desired quadratic term.  Using a mass term merely quadratic in the lapse made it easy to get the exact (not infinitesimal) result at only second order.  Hence the primary constraint effects a  replacement of $N$ by $N-\xi$ in the canonical Lagrangian.  This is true whether $p$ is first-class as in GR or second-class as in massive gravities.

One can follow Pooley and Wallace more closely by performing a first-class transformation at the level of the field equations instead of the canonical action.  That is a bit easier in one respect---the field equations are at most linear in the lapse and shift in the  massive gravity theories chosen---but  less convenient in other respects than is working with the canonical action.   The goal is a relative gauge freedom proof (somewhat akin to 19th century  relative consistency proofs for non-Euclidean geometry, to the effect that non-Euclidean geometry is consistent if Euclidean geometry is consistent).  Hence it will not be necessary to rehash too much of GR, but  only to show that the innovations of massive gravity do not spoil whatever notion of gauge freedom might be generated by the primary constraints.  The Euler-Lagrange equations are
\begin{eqnarray} 
\frac{\partial \mathcal{L}_c }{\partial N} = - \mathcal{H}_0 + m^2 \sqrt{h} - m^2 N = 0,  \nonumber \\
\frac{\partial \mathcal{L}_c }{\partial \beta^i} = - \mathcal{H}_i + m^2 \beta^i = 0, \nonumber \\
\frac{\partial \mathcal{L}_c }{\partial h_{ij} } - \frac{\partial}{\partial t} \frac{\partial \mathcal{L}_c}{\partial \dot{h}_{ij} } - \partial_k \frac{\partial \mathcal{L}_c }{\partial h_{ij},_k }  + \partial_l \partial_k \frac{\partial \mathcal{L}_c }{\partial h_{ij},_{kl} }  = \nonumber
GR\_piece +  \frac{m^2}{2} N \sqrt{h}h^{ij} - \dot{\pi}^{ij}  = \nonumber \\ GR\_pieces  +  \frac{m^2}{2} N \sqrt{h}h^{ij}=  0, \nonumber  \\
\frac{\partial \mathcal{L}_c }{\partial \pi^{ij} }  - \partial_k \frac{\partial \mathcal{L}_c }{\partial \pi^{ij},_k } =  \dot{h}_{ij}  + \left( \frac{\partial }{\partial \pi^{ij} } - \partial_k  \frac{\partial }{\partial \pi^{ij},_k } \right)(-N \mathcal{H}_0 - \beta^i \mathcal{H}_i) = GR\_terms = 0.
\end{eqnarray}
In a relative gauge freedom proof, one can ignore $GR\_pieces$ and $GR\_terms$ to focus attention on the difference that the graviton mass term makes.
Now we can take a Poisson bracket of $\{ \int d^3x \xi(x) p(x)$ with each of these equations of motion:
$$ \{ \int d^3x \xi(x) p(x), \frac{\partial \mathcal{L}_c }{\partial N}(y) \} = m^2 \xi(y),$$
so 
$$ \frac{\partial \mathcal{L}_c }{\partial N}(y)  \rightarrow \frac{\partial \mathcal{L}_c }{\partial N}(N \rightarrow N - \xi),$$ thus behaving as expected.
$\frac{\partial \mathcal{L}_c }{\partial \beta^i} = - \mathcal{H}_i + m^2 \beta^i = 0$ does not depend on $N$ or any derivative thereof, so trivially it is invariant under the primary constraint transformation.
The crucial equation  
$$\frac{\partial \mathcal{L}_c }{\partial h_{ij} } - \frac{\partial}{\partial t} \frac{\partial \mathcal{L}_c}{\partial \dot{h}_{ij} } - \partial_k \frac{\partial \mathcal{L}_c }{\partial h_{ij},_k }  + \partial_l \partial_k \frac{\partial \mathcal{L}_c }{\partial h_{ij},_{kl} }  =0$$ changes by 
$$ - \frac{m^2}{2} \sqrt{h}h^{ij} \xi(y),$$ 
which is just what is required to replace $N$ by $N-\xi.$  
Finally, $$\frac{\partial \mathcal{L}_c }{\partial \pi^{ij} }  - \partial_k \frac{\partial \mathcal{L}_c }{\partial \pi^{ij},_k } =0$$  is exactly as in GR, so a relative gauge freedom proof has no work left to do.  One sees that for all the equations of motion, just as with Proca's massive electromagnetism above, the primary constraint (here $p(x)=0$, using the momentum conjugate to the lapse $N$) generates just as much of a gauge transformation in the massive theory, where the constraint is second-class, as in the massless theory, where the constraint is first-class.  
Hence the conventional wisdom that no second-class constraint generates a gauge transformation but an isolated first-class constraint typically does, faces a puzzle.


\section{Massive Gravity Quadratic in Shift and the Primary Constraint:  Contravariant Weight $2$ Theory} 

Whereas the previous graviton mass term was chosen because it made the dependence on the lapse $N$ very simple, quadratic with a constant coefficient, this theory makes the dependence on the shift  (or something close enough) very simple.  
The theory in question is the Ogievetsky-Polubarinov theory with $n = 1$ (contravariant) with weight $l =2 $ (given the typical western convention, which is the opposite of theirs with parameter $p = -2$).  
The mass term explicitly is 
$$ \frac{m^2}{32 \pi G} ( 6 \sqrt{-g} - (-g)g^{\mu\nu} \eta_{\mu\nu} -2 ) = \frac{m^2}{32 \pi G} ( 6 N\sqrt{h} - h - N^2 {h} h^{ij} \delta_{ij} + \beta^{i}\beta^{i} h -2 ).
$$
This theory is purely quadratic in the shift vector (with a constant coefficient involving the graviton mass) once one redefines the shift to be a weight $1$ contravariant density $\tilde{\beta}^i$ instead of a contravariant vector $\beta^i$ using $\tilde{\beta}^i = \sqrt{h} \beta^i$.  Re-densitizing in order to make calculations in canonical gravity simpler is a standard technique.  Taking $\tilde{\beta}^i$ to be primitive means that $\beta^i$ is now derived:  $\beta^i = \tilde{\beta}^i/\sqrt{h}.$  One now includes $\tilde{\beta}^i$ among the canonical coordinates, defines a canonical momentum with respect to the velocity of $\tilde{\beta}^i$ (which of course is $0$, a primary constraint), \emph{etc.}

Multiplying the mass term by $16 \pi G$ to fit the geometric normalization, one gets the massive gravity  Hamiltonian
$$ \mathcal{H}_{ms} = N \mathcal{H}_0 + \tilde{\beta}^i \mathcal{H}_i/\sqrt{h} - \frac{m^2}{2}(6N\sqrt{h} - h - N^2 h h^{ij} \delta_{ij} + \tilde{\beta}^i \tilde{\beta}^i - 2) + p\dot{N} + \utilde{p}_i \dot{\tilde{\beta}}.$$ 
The canonical Lagrangian is therefore 
$$ \mathcal{L}_c = \pi^{ij} \dot{h}_{ij} - N \mathcal{H}_0 - \tilde{\beta}^i \mathcal{H}_i/\sqrt{h} + \frac{m^2}{2}(6N\sqrt{h} - h - N^2 h h^{ij} \delta_{ij} + \tilde{\beta}^i \tilde{\beta}^i - 2).$$ 
One can now investigate what the smeared primary constraint $$\int d^3x \tilde{\xi}^i(x) \utilde{p}_i(x)$$ does to the canonical Lagrangian density; here $ \utilde{p}_i(x)$ is the momentum conjugate to the densitized shift $\tilde{\beta}^i.$  The tilde below the letter in $\utilde{p}_i(x)$ is intended to show that $ \utilde{p}_i(x)$, which is of weight $0$ (a covector), has one lower density weight than one usually expects of a canonical momentum.  The change in the action from the smeared primary constraint is 
\begin{eqnarray*}
\delta \mathcal{L}_c = \{ \int d^3x \tilde{\xi}^i(x) \utilde{p}_i(x), \mathcal{L}_c(y) \} = \{ \int d^3x \tilde{\xi}^i(x) \utilde{p}_i(x),  - \tilde{\beta}^i \mathcal{H}_i/\sqrt{h}(y) + \frac{m^2}{2}  \tilde{\beta}^i \tilde{\beta}^i(y) \}  = \nonumber \\
 - \tilde{\xi}^k ( - \mathcal{H}_k/\sqrt{h} + m^2 \tilde{\beta}^k)(y) =  - \tilde{\xi}^k \frac{\partial \mathcal{L}_c }{\partial \tilde{\beta}^k }(y). 
\end{eqnarray*}
Thus apart from a term quadratic in $\tilde{\xi}^k$, the primary constraint has simply replaced $\tilde{\beta}^k$ with $\tilde{\beta}^k - \tilde{\xi}^k.$  This is not a divergence, hence not a gauge transformation in the uncontroversial strict sense.  To get the quadratic-in-$\tilde{\xi}^k$ term $\frac{m^2}{2} \tilde{\xi}^k \tilde{\xi}^k $, one can use the next (quadratic) term from an exponentiation of the Poisson bracket.  As before, whether the primary constraint $\utilde{p}_k$ is first-class or second-class has played \emph{no role} in the derivation; one gets the same replacement of $\tilde{\beta}^k$ with $\tilde{\beta}^k - \tilde{\xi}^k$ in non-gauge purely second-class massive spin $2$ gravity as in gauge purely first-class massless spin $2$ gravity, that is GR. %

The Euler-Lagrange equations require a bit of care because now all partial derivatives with respect to variables besides $\tilde{\beta}^i$ are taken with  $\tilde{\beta}^i$ held constant, not $\beta^i.$  This will give some new terms even for GR.  
\begin{eqnarray} 
\frac{\partial \mathcal{L}_c }{\partial N} = - \mathcal{H}_0 + 3m^2 \sqrt{h} - m^2 N h h^{ij} \delta_{ij}=0, \nonumber \\
\frac{\partial \mathcal{L}_c }{\partial \tilde{\beta}^i } =  - \frac{ \mathcal{H}_i }{\sqrt{h} } + m^2 \tilde{\beta}^i=0, \nonumber \\
\frac{\partial \mathcal{L}_c }{\partial h_{ij} } -   \frac{\partial}{\partial t} \frac{\partial \mathcal{L}_c}{\partial \dot{h}_{ij} } - \partial_k \frac{\partial \mathcal{L}_c }{\partial h_{ij},_k }  + \partial_l \partial_k \frac{\partial \mathcal{L}_c }{\partial h_{ij},_{kl} }  = \nonumber \\
\left( \frac{\partial  }{\partial h_{ij} }   - \partial_k \frac{\partial  }{\partial h_{ij},_k }  + \partial_l \partial_k \frac{\partial  }{\partial h_{ij},_{kl} } \right) ( - N \mathcal{H}_0  - \tilde{\beta}^q \frac{ \mathcal{H}_q }{ \sqrt{h} } + \frac{m^2}{2}(6 N \sqrt{h} - h - N^2 h h^{ij} \delta_{ij} + \tilde{\beta}^i \tilde{\beta}^i - 2 ) )  - \dot{\pi}^{ij} = \nonumber \\
 usual\_GR\_bit + \frac{1}{2} h^{ij} \tilde{\beta}^k \mathcal{H}_k/\sqrt{h} - \dot{\pi}^{ij} + \frac{3m^2}{2} N \sqrt{h} h^{ij} - \frac{m^2}{2} h h^{ij} - \frac{m^2}{2} N^2 h \delta_{kl} (h^{ij} h^{kl} - h^{ik} h^{lj}) = 0,\nonumber \\
\frac{ \partial \mathcal{L}_c }{\partial \pi^{ij} } - \partial_k \frac{\partial \mathcal{L}_c }{\partial \pi^{ij},_k } = as\_in\_GR = 0.
\end{eqnarray}
So using the weighted shift has produced a new term proportional to $\mathcal{H}_i,$ which in vacuum GR is the momentum constraint, added to the evolution equations, which is fine.  (Given the graviton mass term, the modified momentum constraint will also involve an $m^2$ term, as appears above.)  The evolution equations also have new pieces from the graviton mass term.  The canonical momentum conjugate to the spatial metric has the same relation to the spatial metric's velocity as in GR.

It is also useful to apply the primary constraint transformation to the equations of motion, much as Pooley \& Wallace did and as was done to the other massive gravity theory considered explicitly above.  One obtains
\begin{eqnarray} 
\delta \frac{\partial \mathcal{L}_c }{\partial N} = \{ \int d^3x \tilde{\xi}^k(x) \utilde{p}_i(x), \frac{\partial \mathcal{L}_c }{\partial N }(y) \} = \tilde{\xi}^k \frac{\partial}{\partial \tilde{\beta}^k } ( - \mathcal{H}_0 + 3m^2 \sqrt{h} - m^2 N h h^{ij} \delta_{ij} )= 0, \nonumber \\
\delta \frac{\partial \mathcal{L}_c }{\partial \tilde{\beta}^i }(y) = \{ \int d^3x \tilde{\xi}^k(x) \utilde{p}(x), \frac{\partial \mathcal{L}_c }{\partial \tilde{\beta}^i } (y) \}= - m^2 \tilde{\xi}^i, \nonumber \\
\delta \left( \frac{\partial \mathcal{L}_c }{\partial h_{ij} } \ldots \right)(y) =  - \frac{1}{2} \tilde{\xi}^k h^{ij} \frac{\mathcal{H}_k }{\sqrt{h} }, \nonumber \\
\delta \left( \frac{ \partial \mathcal{L}_c }{\partial \pi^{ij} }  \ldots \right) = usual\_GR.
\end{eqnarray} 
In all cases, including the GR term that arises from  using a weighted shift $\tilde{\beta}^i,$ one simply replaces  $\tilde{\beta}^i$ by  $\tilde{\beta}^i - \tilde{\xi}^i.$
So the primary constraint transformation generates, at the level of the equations of motion, just as much of a gauge transformation in this massive theory of gravity (where one expects no gauge freedom) as it does in GR.

It is not difficult to show that the constraints for both these massive gravity theories are indeed second-class; a somewhat more general calculation been done previously \cite{PittsQG05}.  
The Poisson brackets among the primary constraints trivially are $0$ because they are pure momenta.  The Poisson brackets among the secondary constraints turn out not to contribute to the determinant, so only the brackets between primary and secondary constraints are needed.  Those involve differentiating the Hamiltonian with respect to the lapse (twice), the shift (or some relative thereof) twice, or the lapse once and the shift (or relative thereof) once.  In the theories studied in this paper, the mixed partial derivatives vanish because no term has both $N$ and $\beta^i.$  For more general theories \cite{OP,MassiveGravity1,MassiveGravity2,MassiveGravity3}, one would expect the lapse  and shift to appear in the same term; one might explore whether field redefinitions, such as a shift-replacement that includes some lapse (not just the spatial metric) can be of use. 

In both massive gravity theories it turns out that at least some of the second-class primary constraints generate gauge transformations just as much as the corresponding first-class primaries do.  But if second-class constraints do not generate gauge transformations, it is unclear why  first-class primaries individually do either.  While the point was evident already after the Proca calculation, doing explicit calculations for massive gravity makes the point more vivid.  It also shows that calculations are not necessarily  difficult due to  nonlinearities in massive gravity, if one chooses mass terms and field definitions judiciously. In fact the calculations can be done far more generally, applying both $p$ and $p_i$ to arbitrary mass terms and for finite smearing functions, if one adapts a more abstract notation, as will now appear.


\section{Finite Translation for General Mass Terms Using Primary Constraints}

The results above do not in fact depend on the careful choice of mass terms to be purely quadratic in either the lapse or the shift (density); such restrictions were for convenience.  One can now show how exponentiating the smeared primary constraint $p$ conjugate to $N$ generates a translation of $N$ by $-\xi$, for $\xi$ finite, for arbitrary mass terms.  In cases where one gets a binomial series expansion rather than a polynomial, one should have $| \frac{\xi}{N}| <1$.  When exponentiating the smeared primary constraint $\int d^3x \xi(x) p(x),$ dependence on phase space quantities unrelated to the lapse is not important, so let us write the canonical Lagrangian as
$$ \mathcal{L}_c = \sum_j c_j N^j.$$   This expression can accommodate power series in $N$, but can also accommodate negative or (with some abuse of notation) even fractional powers, or sums of these. Non-integral and even irrational powers are needed, or at least very helpful, because non-integral and even irrational density weights make sense and lead to distinct massive gravity theories \cite{OP}, in contrast to massive scalar gravities \cite{PittsScalar}, where density weight and powers of the metric cover the same ground twice. 
It would be possible to express these theories using other variables \cite{MassiveGravity3}, but it would be difficult to think them up in the first place. 
 One works term by term, and works out the expansion iteratively until one sees the expected polynomial for $c_j (N-\xi)^j$ or binomial series expansion for $c_j N^j (1 - \frac{\xi}{N})^j$ emerge. As one term is computed, it provides the material for the next term, which then condenses out of the ellipsis to be included explicitly.  
\begin{eqnarray} 
e^{  \{ \int d^3x \xi(x) p(x),   }   c_j(y) N^j(y) \} = c_j N^j(y) +  \{ \int d^3x \xi(x) p(x), c_j(y) N^j(y) \} + \ldots  = \nonumber \\
 c_j N^j(y)  - j c_j N^{j-1} \xi(y) + \frac{1}{2} \{ \int d^3x \xi(x) p(x), - j c_j(y) N^{j-1}(y) \xi(y) \} + \ldots = \nonumber \\
c_j N^j(y)  - j c_j N^{j-1} \xi(y) + \frac{j(j-1)}{2} c_j N^{j-2} \xi^2(y) + \frac{1}{3!} j(j-1) c_j(y) \xi^2(y) \int d^3x \xi(x) \{ p(x), N^{j-2}(y) \} + \ldots = \nonumber \\
c_j N^j(y)  - j c_j N^{j-1} \xi(y) + \frac{j(j-1)}{2} c_j N^{j-2} \xi^2(y) - \frac{1}{3!} j(j-1)(j-2) c_j(y) \xi^3(y) N^{j-3}(y)  + \ldots \nonumber
\end{eqnarray}
If $j$ is a whole number, then this expression terminates.  If not, whether $j$ is negative or nonintegral, the expression gives the binomial series expansion for $c_j N^j (1 - \frac{\xi}{N})^j$.  Then $\xi$ can be finite but not too large.  Once more, whether $p(x)$ is first-class or second-class played \emph{no role} in this derivation.  Hence one sees that, for any massive gravity theory---indeed one is not even restricted to the Ogievetsky-Polubarinov theories or even to Poincar\'{e}-covariant theories---the primary  constraint $p(x)$ conjugate to $N$ generates  as much of a gauge transformation for massive gravity (which should be none) as it does for GR.  

 There is no principial reason that matters should differ for the primary constraint $p_i$ (or $\tilde{p}_i$ or the like, depending on one's choice of variables) conjugate to the shift (density), although the festival of indices could get unattractive.  Hence a somewhat more capacious notation is needed, with an index for indices to accommodate  the fact that different terms have different numbers of indices.  Also it is not clear what a non-whole power of a vector would be (especially if one does not expect spinors to appear), so a whole number $n$ of factors of the shift (or close relative thereof) can be assumed.  A  term takes the form
$$ c_{j_1 \ldots j_n}(y) \beta^{j_1}(y) \ldots \beta^{j_n}(y),$$
where $ c_{j_1\ldots j_n}(y)$ is totally symmetric. A general mass term is some sum of such terms, as far as its dependence on the shift is concerned.  
 The result is 
\begin{eqnarray} 
e^{ \{ \int d^x \xi^i(x) p_i(x), }    c_{j_1 \ldots j_n}(y) \beta^{j_1}(y) \ldots \beta^{j_n}(y) \} = c_{j_1 \ldots j_n}(y) \beta^{j_1}(y) \ldots \beta^{j_n}(y)  - n c_{j_1 \ldots j_n}(y) \xi^{j_1}(y) \beta^{j_2}(y) \ldots \beta^{j_n}(y) \nonumber \\
 + \frac{n(n-1)}{2} c_{j_1 \ldots j_n}(y) \xi^{j_1} \xi^{j_2} \beta^{j_3}(y) \ldots \beta^{j_n}(y)  - \frac{n(n-1)(n-2)}{6} c_{j_1 \ldots j_n}(y) \xi^{j_1} \xi^{j_2} \xi^{j_3} \beta^{j_4}(y) \ldots \beta^{j_n}(y) + \ldots,
\end{eqnarray}
which terminates eventually and is simply $$c_{j_1 \ldots j_{n}} (\beta^{j_1} - \xi^{j_1})\ldots (\beta^{j_n} - \xi^{j_n}).$$ Hence the primary second-class constraint related to the shift also generates just as much a gauge transformation as does the corresponding primary first-class constraint in GR.  
These massive electromagnetic and gravitational theories share with their massless gauge relatives the incompleteness of the Legendre transformation, but the massive theories do not naturally have indeterminism due to  arbitrary functions in the solutions of the equations of motion.  %


\section{Quantization?}

While the separate first-class constraints doctrine and extended Hamiltonian face a puzzle classically, a compelling quantum motivation is not  excluded.  Henneaux and Teitelboim indeed offer a quantum motivation as one reason for postulating that first-class constraints typically generate gauge transformations \cite[p. 18]{HenneauxTeitelboim}.  Quantization was a major motivation when the separate first-class constraints doctrine seems to have made its first appearance in the work of Bergmann and collaborators \cite{BergmannSchiller,BergmannObservables}.  On the other hand, the claim that key use is made of the separate first-class constraints doctrine is disputed by Pons in a section reflecting on why the limitations of  Dirac's work have not caused much trouble \cite{PonsDirac}.  Clearly the truth of the matter is important for the purposes for which one studies gauge theories.  


\section{Declaration of Interests}

The author declares that he has no known competing financial interests or personal relationships that could have appeared to influence the work reported in this paper.


\section{Acknowledgments}

The author thanks Prof. Dr. Claus Kiefer for encouragement.  

Supported by National Science Foundation (USA) grant \#1734402 and by John Templeton Foundation grant \#60745.



 \end{document}